\documentclass{article}
\newtheorem{lemma}{Lemma}[section]

\def\qed{\ \hfill\fbox{}}

\title{{\bf Coloring Artemis graphs}}

\author{%
Benjamin L\'ev\^eque\thanks{Laboratoire Leibniz-IMAG, 46 avenue
F\'{e}lix Viallet, 38031~Grenoble~Cedex, France.
benjamin.leveque@imag.fr. Supported by Ecole Normale Sup\'erieure de
Lyon.}%
\and%
Fr\'ed\'eric Maffray\thanks{C.N.R.S., Laboratoire Leibniz-IMAG, 46
avenue F\'{e}lix Viallet, 38031~Grenoble~Cedex, France.
frederic.maffray@imag.fr}%
\and%
Bruce Reed\thanks{School of Computer Science, McGill University, 3480
University, Montreal, Quebec, Canada H3A 2A7}%
\and%
Nicolas Trotignon\thanks{Laboratoire Leibniz-IMAG, 46 avenue F\'{e}lix
Viallet, 38031~Grenoble~Cedex, France.  nicolas.trotignon@imag.fr.
Supported by Universit\'e Pierre Mend\`es France.}}

\begin{document}
\maketitle

\begin{abstract}
We consider the class ${\cal A}$ of graphs that contain no odd hole,
no antihole, and no ``prism'' (a graph consisting of two disjoint
triangles with three disjoint paths between them).  We show that the
coloring algorithm found by the second and fourth author can be
implemented in time $O(n^2m)$ for any graph in ${\cal A}$ with $n$
vertices and $m$ edges, thereby improving on the complexity proposed
in the original paper.
\end{abstract}

\section{Introduction}

We denote by $\chi(G)$ the chromatic number of a graph $G$ and by
$\omega(G)$ the maximum clique size in $G$.  An \emph{even pair} in a
graph $G$ is a pair $\{x,y\}$ of non-adjacent vertices having the
property that every chordless path between them has even length
(number of edges).  Given two vertices $x,y$ in a graph $G$, the
operation of \emph{contracting} them means removing $x$ and $y$ and
adding one vertex with edges to every vertex of $G\setminus \{x,y\}$
that is adjacent in $G$ to at least one of $x,y$; and we denote by
$G/xy$ the graph that results from this operation.  Fonlupt and Uhry
\cite{fonuhr82} proved that \emph{if $\{x,y\}$ is an even pair in a
graph $G$, then $\chi(G/xy)=\chi(G)$ and $\omega(G/xy)=\omega(G)$}.
In particular, given a $\chi(G/xy)$-coloring $c$ of the vertices of
$G/xy$, one can easily obtain a $\chi(G)$-coloring of the vertices of
$G$ by assigning to $x$ and $y$ the color assigned by $c$ to the
contracted vertex and keeping the color of every vertex different from
$x,y$.  This idea is the basis of a conceptually simple coloring
algorithm: as long as the graph has an even pair, contract any such
pair; when there is no even pair find a coloring $c$ of the contracted
graph and, applying the above procedure repeatedly, derive from $c$ a
coloring of the original graph.  In this perspective, a graph $G$ is
called \emph{even-contractile} \cite{ber90} if it can be turned into a
clique by a sequence of contractions of even pairs, and the graph is
called \emph{perfectly contractile} if every induced subgraph of $G$
is even-contractile.  We propose here a fast implementation of the
above algorithm for a class of perfectly contractile graphs studied in
\cite{maftro03}.

A \emph{hole} is a chordless cycle with at least four vertices and an
\emph{antihole} is the complement of a hole.  A \emph{prism} is a
graph that consists of two vertex-disjoint triangles (cliques of size
three) with three vertex-disjoint paths between them, and with no
other edge than those in the two triangles and in the three paths.
Let $\cal A$ be the class of graphs that contain no odd hole, no
antihole of length at least $5$, and no prism (such graphs have also
been called ``Artemis graphs'' \cite{epsbook}).  Maffray and Trotignon
\cite{maftro03} proved Everett and Reed's conjecture
\cite{epsbook,ree93} that every graph in class ${\cal A}$ is perfectly
contractile.  The proof contains an algorithm which, given any graph
$G$ in class ${\cal A}$ with $n$ vertices and $m$ edges, finds an
optimal coloring of the vertices of $G$ in time $O(n^4m)$.  The point
of this note is to show that this coloring algorithm can be
implemented in time $O(n^2m)$.

In a graph $G=(V,E)$, we say that a vertex $u$ \emph{sees} a vertex
$v$ when $u, v$ are adjacent, else we say that $u$ \emph{misses} $v$.
For any $X \subseteq V$, the subgraph induced by $X$ is denoted by
$G[X]$, and $N(X)$ denotes the set of vertices of $V \setminus X$ that
see at least one vertex of $X$.  A vertex of $V\setminus X$ is called
\emph{$X$-complete} if it sees every vertex of $X$; and $C(X)$ denotes
the set of $X$-complete vertices of $V\setminus X$.  The complementary
graph of $G$ is denoted by $\overline{G}$.  The length of a path is
the number of its edges.  An edge between two vertices that are not
consecutive along the path is a \emph{chord}, and a path that has no
chord is \emph{chordless}.  A vertex is \emph{simplicial} if its
neighbours are pairwise adjacent.

\section{The method}

We recall the method from \cite{maftro03}.  An even pair $\{a,b\}$ in
a graph $G$ is called \emph{special} if the graph $G/ab$ contains no
prism.   
\begin{lemma}[\cite{epsbook,maftro03}]
\label{lem:specep}
If $G$ is in class ${\cal A}$ and $\{a,b\}$ is a special even pair of
$G$, then $G/ab$ is in class ${\cal A}$.
\end{lemma}

The proof from \cite{maftro03} consists in finding a special even pair
and contracting it.  Since Lemma~\ref{lem:specep} ensures that the
contracted graph is still in ${\cal A}$, the algorithm can be iterated
until the graph is a clique.  Actually we will stop when the graph is
a dsjoint union of cliques, which can be colored optimally by the
greedy method.  Let us now recall how a special even pair is found
when the graph is not a disjoint union of cliques.

A non-empty subset $T\subseteq V$ is called \emph{interesting} if
$\overline{G}[T]$ is connected (in short we will say that $T$ is
co-connected) and $G[C(T)]$ is not a clique (so $|C(T)|\ge 2$ since we
view the empty set as a clique).  An interesting set is \emph{maximal}
if it is not strictly included in another interesting set.  A
\emph{$T$-outer path} is a chordless path whose two endvertices are in
$C(T)$ and whose interior vertices are all in $V\setminus (T\cup
C(T))$.  A $T$-outer path $P$ is \emph{minimal} if there is no
$T$-outer path whose interior is strictly contained in the interior of
$P$.  The search for a special even pair considers three cases: (1)
when the graph has no interesting set; (2) when a maximal interesting
set $T$ of $G$ has no $T$-outer path; (3) when a maximal interesting
set $T$ of $G$ has a $T$-outer path.  These three cases correspond to
the following three lemmas.

\begin{lemma}[\cite{maftro03}]\label{lem:int0}
For any graph $G$ the following conditions are equivalent: \\
(1) $G$ has no interesting set,\\
(2) Every vertex of $G$ is simplicial,\\
(3) $G$ is a disjoint union of cliques.\\
Moreover, if $G$ is not a disjoint union of cliques then every
non-simplicial vertex forms an interesting set.
\end{lemma}

\begin{lemma}[\cite{maftro03}]\label{lem:noouter}
Let $G$ be a graph in ${\cal A}$ that contains an interesting set, and
let $T$ be any maximal interesting set in $G$.  If $T$ has no
$T$-outer path, then every special even pair of the subgraph $G[C(T)]$
is a special even pair of $G$.
\end{lemma}

When a maximal interesting set $T$ has a $T$-outer path, we let
$\alpha z_1 \cdots z_p \beta$ be a minimal $T$-outer path and we
define sets:
\begin{eqnarray*}
A&=&\{v\in C(T)\mid vz_1\in E, vz_i\not\in E \ (i=2,\ldots, p)\},\\
B&=&\{v\in C(T)\mid vz_p\in E, vz_i\not\in E \ (i=1,\ldots, p-1)\}.
\end{eqnarray*}
Define a relation $<_A$ on $A$ by setting $u<_A u'$ if and only if $u,
u'\in A$ and there exists an odd chordless path from $u$ to a vertex
of $B$ such that $u'$ is the second vertex of that path (where $u$ is
the first vertex).  Likewise define a relation $<_B$ on $B$ by setting
$v<_B v'$ if and only if $v, v'\in B$ and there exists an odd
chordless path from $v$ to a vertex of $A$ such that $v'$ is the
second vertex of that path.

\begin{lemma}[\cite{maftro03}]
\label{lem:order}
When $A, B$ and $<_A, <_B$ are defined as above they satisfy:\\
(1) The sets $A$ and $B$ are non-empty cliques with no edge between
them.\\
(2) If $P= uu'\cdots v'v$ is a chordless odd path with $u\in A$ and
$v\in B$, then either $u'\in A$ or $v'\in B$ holds.\\
(3) The relation $<_A$ is a strict partial order on $A$.  The relation
$<_B$ is a strict partial order on $B$.\\
(4) If $a$ is any maximal vertex of $<_A$ and $b$ is any maximal
vertex of $<_B$, then $\{a,b\}$ is a special even pair of $G$.
\end{lemma}

\begin{lemma}\label{lem:cpab}
Let $T$ be a maximal interesting set in a graph $G$ and $a,b$ be any
two non-adjacent vertices in $C(T)$.  Let $C'(T)$ be the set of
$T$-complete vertices in $G/ab$.  If $C'(T)$ is not a clique then $T$
is a maximal interesting set in $G/ab$.
\end{lemma}
The proof is easy and we omit it.

We find a special even pair as follows: first an algorithm finds a
maximal interesting set $T$ in $G$.  Then a second algorithm finds a
special even pair in $C(T)$, on the basis of Lemmas~\ref{lem:noouter}
and~\ref{lem:order}, and contracts it; this second algorithm is
iterated as long as the set $C(T)$ is not a clique, which is possible
by Lemmas~\ref{lem:specep} and~\ref{lem:cpab}.  When the set $C(T)$
becomes a clique, the first algorithm is called again to find another
maximal interesting set.  Since the contraction of an even pair
reduces the number of vertices by $1$, there will be at most $n$
contractions.  So the total complexity is $n$ times the complexity of
finding a special even pair.  We will see in the next sections that a
special even pair can be found in time $O(nm)$, so the total
complexity of the coloring algorithm is $O(n^2m)$.


\section{Finding a maximal interesting set}
\label{sec:findint}


\begin{lemma}\label{lem:maxint}
Let $T$ be an interesting set in a graph $G$.  If there is a vertex
$u\in V\setminus(T\cup C(T))$ such that $N(u)\cap C(T)$ is not a
clique then $T\cup\{u\}$ is an interesting set.  If there is no such
vertex then $T$ is a maximal interesting set.
\end{lemma}
The proof is easy and we omit it.


\begin{quote}

{\bf Algorithm Find{\_}Interesting}
  
{\it Input:} A graph $G$.

{\it Output:} Either a maximal interesting set $T$ of $G$ or the
answer ``$G$ is a disjoint union of cliques''.

{\it Method:}
 
{\it Step~1: Looking for a non-simplicial vertex $t$.}

Compute the components of $G$.  If every vertex has degree equal to
the size of its component minus $1$, return the answer ``$G$ is a
disjoint union of cliques'' and stop.  Else, consider a vertex $u$
whose degree is strictly less than the size of its component minus
$1$.  Perform a breadth-first search from $u$, let $v$ be any vertex
at distance $2$ from $u$, and let $t$ be the parent of $v$ in the
search.

{\it Step~2: Building $T$ from $t$.}

Set $T:=\{t\}$, $C:=N(t)$, $U:=V\setminus (T\cup C)$, $Z:=\emptyset$.

While there exists a vertex $u\in U$ do:\\
If $N(u) \cap C$ is a clique, move $u$ from $U$ to $Z$.  \\
If $N(u) \cap C$ is not a clique, move $u$ from $U$ to $T$ and move
every vertex of $C\setminus N(u)$ from $C$ to $U$. 

Return the set $T$ and stop.

\end{quote}


\begin{lemma}
Algorithm Find{\_}Interesting is correct.
\end{lemma}
\emph{Proof.} Clearly, Step~1 of the algorithm is correct.  At the
beginning of Step~2 the set $T$ is interesting and $C$ is equal to the
set of $T$-complete vertices and is not a clique.  The definition of
Step~2 implies that these properties remain true throughout, and
Lemma~\ref{lem:maxint} ensures that when Step~2 terminates the set $T$
is a maximal interesting set.  \qed

\begin{lemma}\label{lem:Ofindint}
The complexity of Algorithm Find{\_}Interesting is $O(\max\{n+m,$
$m(n-k)\})$ where $k$ is the number of vertices in $C(T)$ for the
output set $T$ (if no set $T$ is output, we consider $k=n$ and the
complexity is $O(n+m)$).
\end{lemma}
\emph{Proof.} Step~1 takes time $O(n+m)$ steps.  In Step~2, a vertex
can only move from $C$ to $U$ or from $U$ to $Z$ or to $T$.  So the
sets $T$ and $Z$ can only increase and the sets $U$ and $C$ can only
decrease.  Deciding whether $N(u) \cap C$ is a clique takes time
$O(m)$, and updating $C$ takes time $O(\deg (t))$ since $C$ can only
decrease from its initial value $N(t)$.  Thus, each iteration of the
while loop takes time $O(m)$.  A vertex plays the role of $u$ at most
once, and the $k$ vertices that are in $C$ when the algorithm stops
have never played such a role.  So there are at most $n-k$ iterations
of the while loop.  \qed

\section{Looking for an outer path}

\begin{lemma}\label{lem:int}
An interesting set $T$ has an outer path if and only if there exists a
component $R$ of $V\setminus (T\cup C(T))$ such that $N(R)\cap C(T)$
is not a clique.
\end{lemma}
The proof is easy and we omit it.

\begin{lemma}[\cite{maftro03}]
	\label{lem:outer4}
Let $G$ be a graph in ${\cal A}$ that contains an interesting set, and
let $T$ be any maximal interesting set in $G$.  Then every $T$-outer
path has length even and at least $4$.
\end{lemma}

Given two disjoints subsets $X, Y\subseteq V$ of a graph $G$, we call
\emph{breadth first search (BFS) from $X$ to $Y$ in $G$} any breadth
first search such that (a) the vertices of $X$ form the root level,
and (b) the vertices of $Y$ may only appear as leaves in the search
tree.  Points (a) and (b) can be implemented as in the usual form of
BFS by using a queue from which we get the next vertex to be scanned,
the only modification being that we put all vertices of $X$ in the
queue at the start of the search and we never put any vertex of $Y$ in
the queue.   

\begin{quote}
{\bf Algorithm Find{\_}Outer{\_}Path}

{\it Input:} A graph $G$ and a maximal interesting set $T$ of $G$.

{\it Output:} Either a minimal $T$-outer path or the answer ``$G$ has
no $T$-outer path''.

{\it Method:} Initially all vertices of $V(G)\setminus (T\cup C(T))$
are unmarked.  \\
{\bf while} there is an unmarked vertex $r$ in $V(G)\setminus (T\cup
C(T))$ do:\\
Start a Breadth-First Search $S$ from $r$ to $C(T)$ in $G\setminus T$,
mark each vertex of $V(S)\setminus C(T)$, and maintain the set $M=V(S)
\cap C(T)$.  \\
When a vertex $x$ is added to $M$, {\bf if} the new $M$ is not a
clique, do:\\
Let $M_x:=M\cap N(x)$.  Perform a BFS from $x$ in the subgraph
$G[S\setminus M_x]$.  Let $y$ be the first vertex of $M\setminus M_x$
that is reached by this search, and let $x$-$v$-$\cdots$-$w$-$y$ be
the path from $x$ to $y$ given by this search.  Return this path and
stop.  {\bf endif}\\
{\bf endwhile}\\
Return the answer ``$G$ has no $T$-outer path'' and stop.
\end{quote}

\begin{lemma}
Algorithm Find\_Outer\_Path is correct.
\end{lemma}
\emph{Proof.} Let $R$ be the component of $V(G)\setminus (T\cup C(T))$
that contains $r$.  The search from $r$ potentially reaches all
vertices of $R$ and of $N(R)\cap C(T)$ and puts the latter into $M$.
If $N(R)\cap C(T)$ is a clique, the search will mark all vertices of
$R$ and continue with a new vertex $r$ from another component of
$V(G)\setminus (T\cup C(T))$, if any.  Lemma~\ref{lem:int} ensures
that the algorithm will correctly return the answer ``$G$ has no
$T$-outer path'' if and only if $G$ has no outer path.  There remains
to show that when the algorithm returns a path, it is a minimal
$T$-outer path.  So let us examine the situation in this case.  For
some component $R$ of $V(G)\setminus (T\cup C(T))$ the set $N(R)\cap
C(T)$ is not a clique, and the search from a vertex $r\in R$ finds the
vertex $x$ on the first time $M$ is no longer clique.  The set
$M\setminus M_x$ is not empty.  The only neighbours of $x$ in
$S\setminus M$ are either its parent in the search or vertices that
are still in the queue, for otherwise $x$ would have been added to $S$
earlier.  So every vertex of $S\setminus M$ that is not adjacent to
$x$ has been scanned before the neighbours of $x$ in $S\setminus M$.

The search from $x$ in $G[S\setminus M_x]$ will potentially reach all
vertices of $M\setminus M_x$.  So the vertex $y$ and the path
$x$-$v$-$\cdots$-$w$-$y$ exist.  Let us rewrite this path as
$P=x$-$z_1$-$\cdots$-$z_p$-$y$, with $z_1=v$ and $z_p=w$, and write
$Z=\{z_1, \ldots, z_p\}$.  To show that $P$ is a $T$-outer path,
suppose on the contrary that some element $z_i$ of $Z$ is in $C(T)$
and let $i$ be the smallest such integer ($1\le i\le p$).  We have
$i\ge 2$ because $z_1=v$ which is in $R$; but then $z_i$ contradicts
the definition of $y$.  So all of $z_1, \ldots, z_p$ are in $R$, which
means that $P$ is a $T$-outer path.  To prove the minimality of $P$,
suppose on the contrary that there exists a $T$-outer path
$x'$-$z_i$-$\cdots$-$z_j$-$y'$ with $1\le i\le j\le p$ and $j-i<p-1$.
By Lemma~\ref{lem:outer4}, $j-i$ is even and at least $2$, so $j>2$.
Vertex $y'$ is in $M\setminus\{x\}$ since $z_j$ has been added to $S$
before $z_1$.  If $i>1$ then $x'$ too is in $M\setminus \{x\}$, since
$z_i$ has been added to $S$ before $z_1$; but then $M\setminus\{x\}$
is not a clique, which contradicts the definition of $x$.  So $i=1$.
If $x$ is adjacent to $y'$ then $x$-$z_1$-$\cdots$-$z_j$-$y'$-$x$ is a
hole of odd length $j-1+3\ge 5$, a contradiction.  So $x$ is not
adjacent to $y'$, but then $x$-$z_1$-$\cdots$-$z_j$-$y'$ is a
chordless path and $y'$ contradicts the definition of $y$.  So $P$ is
a minimal $T$-outer path.  \qed

\begin{lemma}\label{lem:Olm}
The complexity of Algorithm Find\_Outer\_Path is $O(lm)$, where $l$ is
the number of components of $G\setminus(T\cup C(T))$.
\end{lemma}
\emph{Proof.} The search from a vertex $r$ reaches all the vertices of
the component $R$ of $V(G)\setminus (T\cup C(T))$ that contains $r$
and the vertices of $N(R)\cap C(T)$, and only them.  Moreover these
vertices are scanned only once during this search.  The search from
$x$ reaches vertices of $R$ a second time.  Thus vertices of $R$ are
scanned at most twice.  In order to check whether $M$ is a clique, we
use a counter for each vertex $u$ of $C(T)$, which counts the number
of neighbours of $u$ in $M$.  Whenever a new vertex $u$ is added to
$M$, we check if the counter of $u$ is equal to $|M|$, and we scan $u$
to increase by $1$ the counter of its neighbours in $C(T)$.  Thus the
complexity for one component $R$ is $O(m(R))$, where $m(R)$ is the
number of edges in the subgraph induced by $R\cup (N(R)\cap C(T))$.
However, a vertex $u$ of $C(T)$ may be scanned several times,
depending on the number of sets of the type $N(R)\cap C(T)$ that
contain it.  So the total complexity is $O(lm)$.  \qed

\

When the set $T$ has no $T$-outer path, Lemma~\ref{lem:noouter} says
that we need to continue the search recursively in the subgraph
$G[C(T)]$.  Thus we may have to find a maximal interesting set $T_1$
of $G$, then (putting $C_1=C(T_1)$) find a maximal interesting set
$T_2$ of $G[C_1]$, then (putting $C_2=C(T_2)\cap C_1$) find a maximal
interesting set $T_3$ of $G[C_2]$, up to (putting $C_{q-1}= C(T_{q-1})
\cap C_{q-2}$) a maximal interesting set $T_q$ of $G[C_{q-1}]$ such
that either there is a $T_q$-outer path in $G[C_{q-1}]$ or $G[C(T_q)
\cap C_{q-1}]$ is a disjoint union of cliques.  Let us analyze the
complexity of this procedure.

For $i=1, \ldots, q$, put $n_i= |C_i|$ and $m_i=|E(G[C_i])|$, and put
$n_0=n$ and $m_0=m$.  By Lemma~\ref{lem:Ofindint}, the total
complexity of finding interesting sets over all the recursive calls is
$O(\Sigma_{i=1}^{q-1} m_{i-1} (n_{i-1}- n_{i}) + \max\{n_{q-1}+
m_{q-1},$ $m_{q-1} (n_{q-1}- n_{q})\})$ $=O(\max\{n+m,$ $mn)\}$.

For $i=1, \ldots, q$, let $l_i$ be the number of components of
$G[C_{i-1}\setminus (T_i\cup C(T_i))]$.  Observe that all these
components (over all $i=1, \ldots, q$) are pairwise disjoint, so
$l_1+\cdots+l_q\le n$.  By Lemma~\ref{lem:Olm}, the total complexity,
over all recursive calls, of finding outer paths is $O(\Sigma_{i=1}^q
l_im)$ $=O(nm)$.  

So the total complexity of this recursive procedure is $O(nm)$.

\section{Finding a special even pair}

\begin{quote}
	{\bf Algorithm Find\_Even\_Pair}

{\it Input:} A graph $G$, a maximal interesting set $T$ and the
minimal $T$-outer path $x$-$v$-$\cdots$-$w$-$y$ given by Algorithm
Find\_Outer\_Path.

{\it Output:} A special even pair of $G$

{\it Method:} \\
1.  Set $A:=(N(v)\cap C(T)) \setminus N(y)$ and $B:=(N(w) \cap C(T))
\setminus N(x)$.\\
2.  Perform a BFS from $B$ to $N(A)$ in $G\setminus (T\cup A)$ and
call $K$ the set of vertices of $N(A)$ that are reached by this
search.\\
3.  Perform a BFS from $A$ to $N(B)$ in $G\setminus (T\cup B)$ and
call $L$ the set of vertices of $N(B)$ that are reached by this
search.\\
4.  Let $a$ be a vertex of $A$ that sees all of $K$.\\
5.  Let $b$ be a vertex of $B$ that sees all of $L$.\\
6.  Return the pair $\{a,b\}$.
\end{quote}

\begin{lemma}
The preceding algorithm returns a pair of vertices $\{a,b\}$ that is a
special even pair of $G$.
\end{lemma}
\emph{Proof.} Let us rewrite the path $x$-$v$-$\cdots$-$w$-$y$ as
$P=x$-$z_1$-$\cdots$-$z_p$-$y$, with $z_1=v$ and $z_p=w$, and write
$Z=\{z_1, \ldots, z_p\}$.  Define sets $A'=\{u\in C(T) \mid uz_1 \in
E(G), uz_i \notin E(G) (i=2,\ldots, n) \}$ and $B'=\{u\in C(T) \mid
uz_n \in E(G), uz_i \notin E(G) (i=1,\ldots, n-1) \}$.  These are the
sets mentioned in Lemma~\ref{lem:order}.  We claim that the sets $A,
B$ defined in the algorithm satisfy $A=A'$ and $B=B'$.  First observe
that $x\in A'$ and $y\in B'$ and that there is no edge $a'b'$ with
$a'\in A'$ and $b'\in B'$, for otherwise $Z\cup\{a', b'\}$ would
induce an odd hole of length $p+2\ge 5$.  This implies $A'\subseteq A$
and $B'\subseteq B$.  Now let $a$ be any vertex of $A$.  Suppose that
$a$ has a neighbour $z_i$ in $Z$ with $i\ge 2$, and let $i$ be the
largest such integer.  By the definition of $A$, vertices $a$ and $y$
are non-neighbours.  Then $a$-$z_i$-$\cdots$-$z_p$-$y$ is a $T$-outer
path, which contradicts the minimality of $P$.  So $a$ has no
neighbour in $Z\setminus\{z_1\}$.  So $A\subseteq A'$.  Similarly
$B\subseteq B'$.  So $A=A'$ and $B=B'$ as claimed.

There remains to show that lines 2--5 of the algorithm correctly
produce maximal elements of $(A, <_A)$ and $(B, <_B)$.  Let $K$ be as
defined by the algorithm.  Let $a^*$ be a maximal vertex for the
relation $<_A$.  Suppose $a^*$ is not adjacent to a vertex $u$ of $K$.
Let $a'\in A\setminus \{a^*\}$ be a neighbour of $u$ ($a'$ exists by
the definition of $K$), and let $Q=q_1$-$\cdots$-$q_k$ be the
chordless path from $q_1\in B$ to $u=q_k$ given by the search tree of
line~2 of the algorithm.  Since $A$ is a clique, $a^*$ and $a'$ are
adjacent.  Suppose a vertex of $A$ is adjacent to a vertex $q_i$ with
$1\le i\le k-1$.  Then $q_i$ is a vertex of $N(A)$ and the search
should not have been continued from $q_i$; but this contradicts the
existence of $q_{i+1}$.  So $a'$ and $a^*$ are not adjacent to any of
$q_1, \ldots, q_{k-1}$.  Let $Q'=q_1$-$\cdots$-$q_k$-$a'$ and
$Q^*=q_1$-$\cdots$-$q_k$-$a'$-$a^*$.  Then $Q'$ and $Q^*$ are
chordless paths.  Lemma~\ref{lem:order} implies that $Q'$ has even
length since none of its interior vertices are in $A\cup B$.  It
follows that $Q^*$ is odd, which implies that $a^*<_A a'$, which
contradicts the choice of $a^*$.  This proves that every maximal
vertex of $<_A$ is adjacent to all of $K$, and consequently that the
vertex $a$ of the algorithm exists.  Conversely, let us prove that any
such $a$ is maximal for the relation $<_A$.  Suppose the contrary.
Then, by the definition of $<_A$, there exists an odd chordless path
$Q''=a$-$a''$-$q$-$\cdots$-$b''$ from $a$ to a vertex $b''\in B$ with
$a''\in A$.  Then $q$ is in $N(A)$ and on a chordless path from $B$,
so $q$ has been reached by the BFS defined on line~2, so $q$ is in
$K$.  But then $a$ is adjacent to $q$, which contradicts the fact that
$Q''$ is chordless.  So $a$ is maximal for the relation $<_A$.  The
proof is similar for $B$: a vertex of $B$ is maximal for the relation
$<_B$ if and only if it is adjacent to all of $L$.  Now
Lemma~\ref{lem:order} implies that the pair $\{a,b\}$ returned by the
algorithm is a special even pair of $G$ and the proof of correctness
is complete.  \qed

\begin{lemma}
The complexity of Algorithm Find\_Even\_Pair is $O(m)$.
\end{lemma}
\emph{Proof.} Determining the sets $A$ and $B$ takes time
$O(d(v)+d(y))$ and $O(d(w)+d(x))$ respectively.  Performing the
breadth-first search from $B$ to $N(A)$ and determining the set $K$
takes time $O(m)$, and similarly for determining the set $L$.
Moreover, each time a vertex is put into $K$ we add $+1$ to a counter
associated to each of its neighbours in $A$.  And we do similarly for
$L$ and $B$.  So finding vertices $a$ and $b$ takes time $O(|A|)$ and
$O(|B|)$ respectively.  \qed

\section{Analogy between interesting sets and handles}

Recall that a graph is \emph{weakly chordal} if $G$ and its
complementary graph contain no hole of length at least $5$.  A
\emph{handle} \cite{hay97a,hay97b} in a graph $G=(V,E)$ is a subset
$H\subset V$, of size at least $2$, such that $G[H]$ is connected,
some component $J \neq H$ of $G\setminus N(H)$ satisfies $N(J) =N(H)$,
and each vertex of $N(H)$ sees at least one vertex of each edge of
$G[H]$.  Any such $J$ is called a \emph{cohandle} of $H$.  Hayward,
Spinrad and Sritharan \cite{hayspi} use handles to obtain a
recognition algorithm for weakly chordal graphs with complexity
$O(m^2)$ and a coloring algorithm for those graphs with complexity
$O(n^3)$.  We observe that there is an analogy between handles and
interesting sets.

\begin{lemma}
\label{thpi}
Let $H$ be a handle of $G$ and $J$ a co-handle of $H$.  Then $J$ is an
interesting set of $\overline{G}$.
\end{lemma}
\emph{Proof.} By the definition of a handle, $G[J]$ is connected.
Moreover, in the graph $\overline{G}$ we have $H\subseteq C(J)$, and
$H$ is connected and $|H|\geq 2$.  So $J$ is an interesting set in
$\overline{G}$.  \qed

\begin{lemma}
Let $T$ be a maximal interesting set in $G$, and let $H$ be a
co-connected component of $G[C(T)]$ of size at least $2$.  Then $H$ is
a handle of $\overline{G}$ and $T$ is a co-handle of $H$.
\end{lemma}
\emph{Proof.} Since $C(T)$ is not a clique in $G$, there exists a
component $H$ of $\overline{G}[C(T)]$ of size at least $2$.  Let
$X=N_{\overline{G}}(H)$.  Since $T$ is connected in $\overline{G}$,
there is a component $T'$ of $\overline{G}\setminus X$ that contains
$T$.  If $T\neq T'$, there is a vertex $u \in V\setminus (X\cup T)$
such that (in $\overline{G}$) $u$ has a neighbour in $T$.  Then (in
$G$) $T\cup\{u\}$ is an interesting set because $T\cup\{u\}$ is
co-connected and $H\subseteq C(T\cup\{u\})$.  This contradicts the
maximality of $T$.  So $T$ is a component of $\overline{G}\setminus
X$.  Now let $Y=N_{\overline{G}}(T)$.  Clearly $Y\subseteq X$.  In $G$
every vertex $x$ of $X$ has a non-neighbour in $H$ and thus is not in
$T\cup C(T)$, and so $x \in Y$.  Therefore $Y = X$.  Finally, suppose
that (in $\overline{G}$) some vertex $x \in X$ misses both vertices of
an edge of $\overline{G}(H)$.  Then (in $G$) the set $T\cup\{x\}$ is
an interesting set strictly larger than $T$, a contradiction.  So $H$
is a handle and $T$ is a co-handle of $H$ in $\overline{G}$.  \qed

The preceding two results show that any maximal interesting set of $G$
gives a handle of $\overline{G}$, but a handle of $\overline{G}$ gives
only an interesting set of $G$, which is not necessarily maximal.
This suggests the following new definition which will strengthen the
correspondence.  A \emph{generalized handle} is a subset $H\subset V$
that contains at least one edge, such that some component $J \neq H$
of $G\setminus N(H)$ satisfies $N(J) = N(H)$, and every vertex of
$N(H)$ sees at least one vertex of each edge of $G[H]$.  Any such $J$
is called a \emph{generalized co-handle} of $H$.  This new definition
of handles still enables us to use ideas from \cite{hayspi}, where the
hypothesis of connectedness of $H$ does not seem to be necessary.  A
handle is a particular type of generalized handle, and the algorithm
\emph{find-handle} of \cite{hayspi} can be modified as follows:

\begin{quote}
{\bf Algorithm Find\_Generalized\_Handle}

Search for a vertex $v$ and an edge $e$ such that $v$ misses $e$\\
{\bf If} no such $v,e$ exist {\bf then} return ``no handle'' and stop {\bf
endif}\\
$J$ $\leftarrow$   component of $G\setminus N(e)$ containing
$v$\\
$H$ $\leftarrow$ $V\setminus (J\cup N(J))$\\
{\bf while} some $v$ in $N(H)$ misses some $e$ in $H$ {\bf do:}\\
$J$ $\leftarrow$   component of $G\setminus N(e)$ containing
$v$\\
$H$ $\leftarrow$ $V\setminus (J\cup N(J))$\\
{\bf endwhile}\\
return $(H,J)$
\end{quote}

\begin{lemma}\label{thm:cohint}
The co-handle produced by this algorithm is a maximal interesting set
in $\overline{G}$.
\end{lemma}
\emph{Proof.} By Lemma~\ref{thpi}, $J$ is an interesting set of
$\overline{G}$.  Suppose that $J$ is not maximal.  So there exists
$j\notin J$ such that $J'=J\cup \{j\} $ is an interesting set of
$\overline{G}$.  Since $G[J']$ is connected, $j$ is in $N(J)$.
However, $N(J) = N(H)$ so $j$ sees at least one vertex of each edge of
$H$.  Thus $V\setminus (J'\cup N(J')) \subset V\setminus (J\cap N(J))
=H$ and so $V\setminus (J'\cup N(J'))$ is a stable set and $J'$ is not
an interesting set, a contradiction.  \qed

Lemma~\ref{thm:cohint} points to an alternative way to find a maximal
interesting set.  However, algorithms to find a handle so far
\cite{hayspi} have not broken the complexity barrier that would make
them better than the one we presented in Section~\ref{sec:findint}.


\begin{thebibliography}{99}

\bibitem{ber90}
M.E.~Bertschi, Perfectly contractile graphs.  \emph{J. Comb.~Th.
B} {50} (1990), 222--230.

\bibitem{epsbook}
H.~Everett, C.M.H.~de~Figueiredo, C.~Linhares~Sales, F.~Maffray,
O.~Porto, B.A.~Reed.  \newblock Even pairs.  In: \emph{Perfect
Graphs}, J.L.~Ram\'{\i}rez-Alfons\'{\i}n and B.A.~Reed, eds., Wiley
Interscience (2001), 67--92.

\bibitem{fonuhr82}
J.~Fonlupt, J.P.~Uhry.  Transformations which preserve perfectness and
$h$-perfectness of graphs.  \emph{Ann.~Disc.  Math.} {16} (1982),
83--85.

\bibitem{hay97a}
R.~Hayward.  Meyniel weakly triangulated graphs I: co-perfect
orderability.  Discrete Applied Mathematics 73 (1997), 199--210.

\bibitem{hay97b}
R.~Hayward.  Meyniel weakly triangulated graphs II: A theorem of
Dirac.  Discrete Applied Mathematics 78 (1997), 283--289.

\bibitem{hayspi}
R.B.~Hayward, J.P.~Spinrad, R. Sritharan.  Weakly chordal graph
algorithms via handles.  Proc.~11th annual ACM-SIAM Symp.~on Discrete
Algorithms, 2000, 42--49.

\bibitem{maftro03}
F.~Maffray, N.~Trotignon.  A class of perfectly contractile graphs.
Leibniz Research Report 67, Grenoble, France. Submitted for publication.
\newline
http://www-leibniz.imag.fr/NEWLEIBNIZ/LesCahiers/2002/Cahier67.



\bibitem{ree93}
B.A.~Reed.  Problem session on parity problems (Public communication).
\emph{DIMACS Workshop on Perfect Graphs}, Princeton Univ., New Jersey,
1993.

\bibitem{tro}
N. Trotignon.  Graphes parfaits: structure et algorithmes.  Doctoral
Thesis, University Joseph Fourier, Grenoble, France, 2004.

\end{thebibliography}
\end{document}